\documentclass[prb,superscriptaddress,twocolumn,showpacs,floatfix]{revtex4}

\usepackage{epsfig,dcolumn,amsmath,latexsym}

\usepackage{subfigure}
\usepackage[normalem]{ulem}
\usepackage{cancel}
\usepackage{multirow}

\def\k233{K$_2$Cr$_3$As$_3$}

\def\k133{KCr$_3$As$_3$}

\begin{document}

\title{Reduced Dimensionality and Magnetic Frustration in KCr$_3$As$_3$}
\preprint{1}

\author{Chao Cao}
 \email[E-mail address: ]{ccao@hznu.edu.cn}
 \affiliation{Condensed Matter Group,
  Department of Physics, Hangzhou Normal University, Hangzhou 310036, P. R. China}

\author{Hao Jiang}
 \affiliation{Department of Physics, Zhejiang University, Hangzhou 310013, P. R. China}

\author{Xiao-Yong Feng}
\affiliation{Condensed Matter Group,
  Department of Physics, Hangzhou Normal University, Hangzhou 310036, P. R. China}

\author{Jianhui Dai}
 \email[E-mail address: ]{daijh@hznu.edu.cn}
 \affiliation{Condensed
Matter Group,
   Department of Physics, Hangzhou Normal University, Hangzhou 310036, P. R. China}

\date{April 18, 2015}

\begin{abstract}

We study the electronic and magnetic structures of the
newly-discovered compound KCr$_3$As$_3$. The non-magnetic state has
five Fermi surface sheets involving respectively three
quasi-one-dimensional and two three-dimensional energy bands.
However, the ground state is magnetic, exhibiting a novel interlayer
antiferromagnetic order where the basic block-spin state of a unit
Cr triangle retains a high spin magnitude. Moreover, its Fermi
surface involves three one-dimensional sheets only, providing
evidence for local moments in this compound due to the reduced
dimensionality. By fitting a twisted spin tube model the magnetic
frustrations caused by local moments are found to be relaxed,
leading to gapless spin excitations. A frustration-induced
transition to the disordered low block-spin state is expected upon
increasing the intralayer exchange interaction.

\end{abstract}

\maketitle

{\it Introduction}.  The interplay between superconductivity and
magnetism or other competing density-waves has been a major theme in
modern condensed matter physics, particularly in transition metal
compounds with both strong electron correlations and reduced
dimensionality. Although the development of long-range orders at
finite temperature is forbidden in strictly one-dimensional system
with short-range
interactions\cite{PhysRev.158.383,PhysRevLett.17.1133}, such
restriction can be lifted for quasi-one-dimensional (Q1D) systems.
Typical examples include the Q1D superconductors
(TMTSF)$_2X$\cite{TMTSF_1}, $M_2$Mo$_6$Se$_6$
($M$=Tl,In)\cite{Brusetti1988181},
Li$_{0.9}$Mo$_6$O$_{17}$\cite{Greenblatt1984671,PhysRevLett.108.187003},
etc. A "universal" phase diagram for the Q1D superconductors
suggests that the superconductivity is in close proximity to the
density-waves\cite{Wilhelm:2001uq,TMT_review}.

The recently discovered Q1D superconductors A$_2$Cr$_3$As$_3$ ( A=K,
Rb, Cs) provide a new setting for understanding the interplay of
unconventional superconductivity and density wave instabilities in
reduced
dimensions.\cite{PhysRevX.5.011013,PhysRevB.91.020506,Tang15} The
building blocks of these compounds consist of Q1D [CrAs]$_\infty$
double-wall nano-tubes intercalated by A$^{+}$ cations. The first
principle calculations using the density functional theory
(DFT)\cite{Jiang14,Wu15} predicted a three-dimensional (3D) Fermi
surface (FS) sheet and two Q1D FS sheets, all contributed mainly by
the Cr-$3d$ electrons. The penetration depth
measurement\cite{Pang15} has evidenced a nodal line in the pairing
state below $T_c$. Meanwhile, the NMR experiment\cite{Zhi15} has
revealed a power law behavior of the spin-lattice relaxation rate,
manifesting the 1D Luttinger liquid feature at the normal state. A
couple of theoretical proposals have been devoted to the possible
pairing symmetries of the superconducting state, concerning the
respective relevance of the Q1D and 3D bands\cite{Zhou15,Wu15b},
while the formation of molecular orbital bands has been detailed in
a twisted Hubbard tube.\cite{Zhong15} These studies suggest possible
competing superconducting and density wave ordering tendencies in
K$_2$Cr$_3$As$_3$. In fact, superconductivity associated with a
magnetic quantum critical point has revealed in the bulk materials
of chromium arsenides\cite{Wei14,Kotegawa14}. The material realization
of density-wave instabilities competing with superconductivity in
the Q1D chromium arsenide compounds should be very interesting and
require more surveys.

Motivated by this issue, we will study the electronic and magnetic
structures of a related compound, KCr$_3$As$_3$.  This 133-type of
chromium arsenide has been recently synthesized
experimentally\cite{Cao133}. It consists of the Q1D
[CrAs]$_{\infty}$ structure similar to K$_2$Cr$_3$As$_3$. The
compound is metallic down to the lowest measured temperature, but no
superconductivity has been observed. Instead, the polycrystal sample
displays a spin glassy feature at low temperatures\cite{Cao133}. In
this paper, we find based on DFT calculations that the electronic
and magnetic structures of \k133\ exhibit new properties in contrast
to the superconducting 233-type compound. The most prominent
features are the lack of three-dimensional Fermi surface sheets and
emergence of the interlayer antiferromagnetic order. These numerical
results indicate coexistence of local moments and itinerant
electrons in the \k133\ compound due to the reduced dimensionality.
Such exotic metallic magnetic phase is understood as due to the
formation of high spin states in each Cr triangles via the orbital
selective Mott transition. A frustration-induced quantum phase
transition to the disordered low block-spin state with a spin gap is
expected by tuning the intralayer exchange to the antiferromagnetic
side. Our study thus reveals a possible interplay between
unconventional superconductivity and magnetic order in this class of
materials.

{\it Calculation Method and Details}.  All reported results are
obtained using the DFT as implemented in the Vienna Abinit
Simulation Package (VASP)\cite{vasp_1,vasp_2}, where the valence
electron-ion interactions are modeled using projected augmented wave
(PAW) method\cite{bloch_paw,vasp_2} and the exchange-correlation
effects are approximated with Perdew-Burke-Enzerhoff flavor of
general gradient approximation (PBE).\cite{PBE_1} To ensure the
convergence of the calculations, a 540 eV energy cutoff to the plane
wave basis and a 6 $\times$ 6 $\times$ 12  $\Gamma$-centered K-mesh
is employed; whereas a 12 $\times$ 12 $\times$ 24 $\Gamma$-centered
K-mesh and tetrahedron method are employed to perform the DOS
calculations.
  \begin{figure}[htp]
    \subfigure[]{
      \includegraphics[width=4cm]{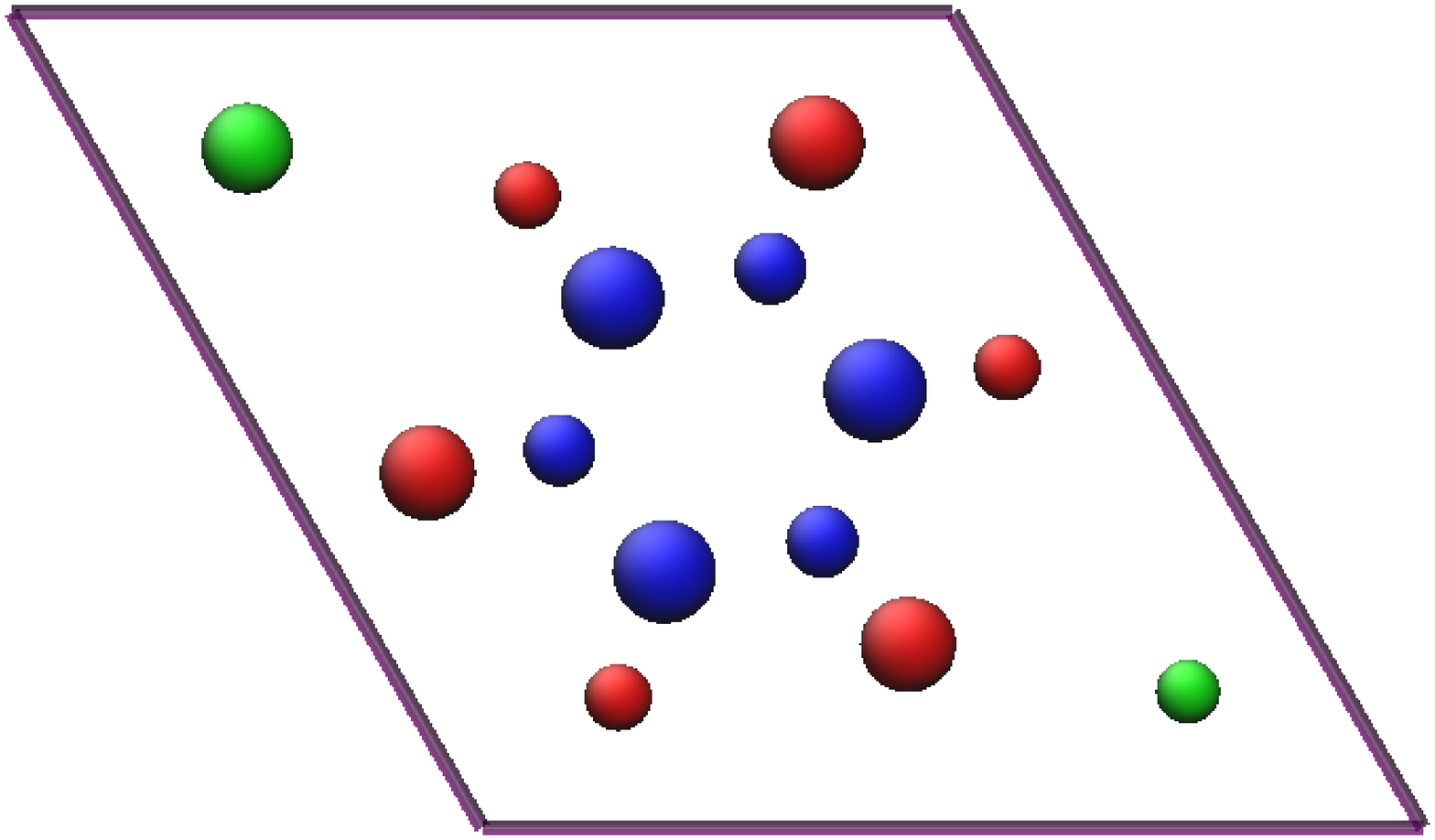}
      \label{fig:geo1}}
    \subfigure[]{
      \includegraphics[width=4cm]{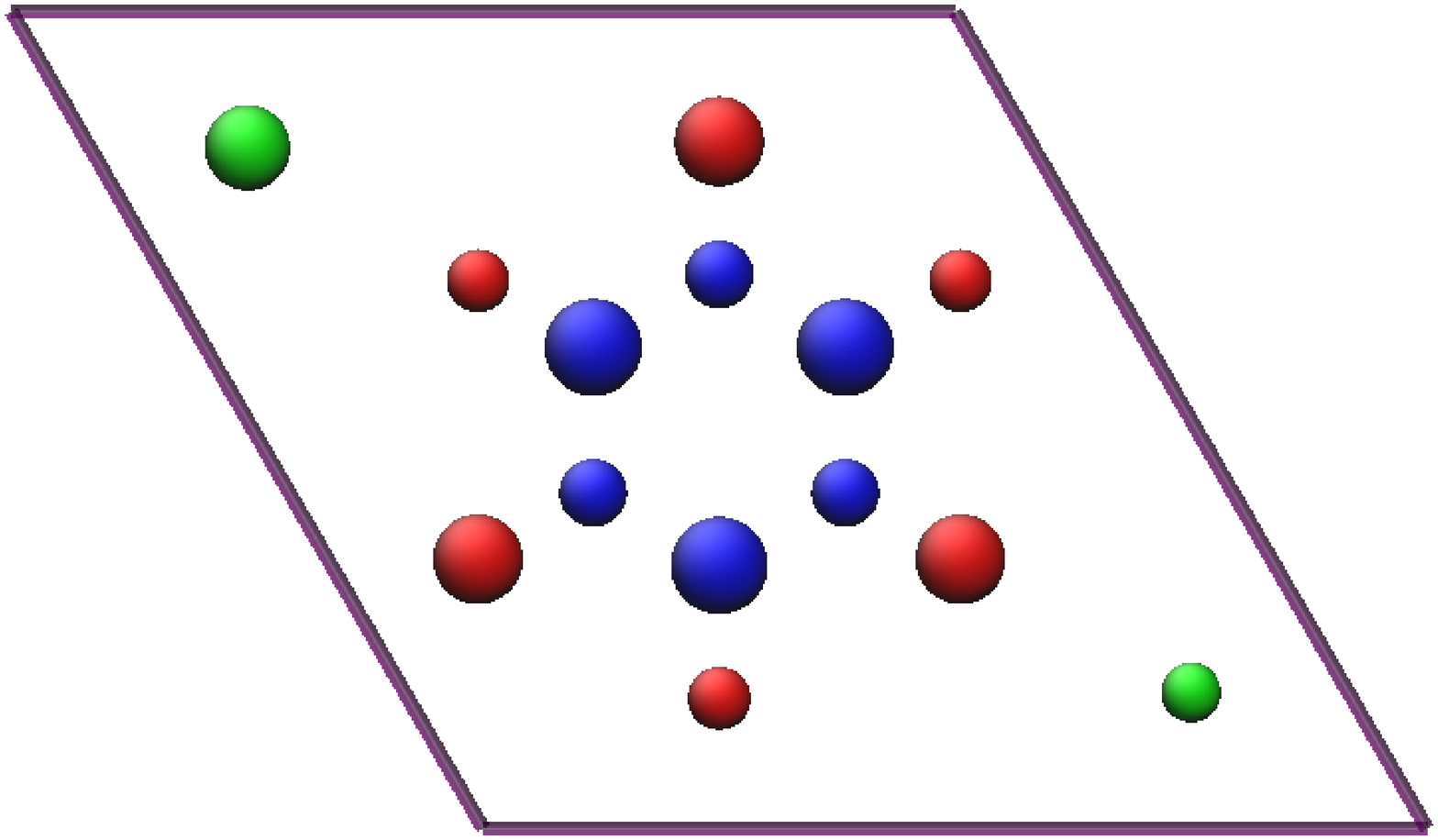}
      \label{fig:geo2}}
    \caption{Crystal structure of \k133. (a) Experimentally determined geometry with space group P6$_3$/m (No. 176); (b) Hypothetic geometry closer to K$_2$Cr$_3$As$_3$ with space group P$\bar{6}$m2 (No. 187). In both panels, the red, blue and green balls represents As, Cr and K, respectively; larger and smaller balls indicate atoms at z=0.75 ($X^{\mathrm{I}}$) and z=0.25 ($X^{\mathrm{II}}$), respectively ($X$=K, Cr, and As). \label{fig:geomag}}
  \end{figure}

Experimentally, the crystal structure of \k133\  shown in FIG.
\ref{fig:geo1}, was determined to be centrosymmetric with space
group P6$_3$/m (No. 176). Unlike the K$_2$Cr$_3$As$_3$ structure,
the Cr$^{\mathrm{I}}$-Cr$^{\mathrm{I}}$ bond lengths are identical
to the Cr$^{\mathrm{II}}$-Cr$^{\mathrm{II}}$ bond lengths in
KCr$_3$As$_3$. Apart from this apparent difference, the [CrAs]$_6$
sub-nanotubes in \k133\ are rotated $\sim 10^{\circ}$ with respect to
their central axis compared to the K$_2$Cr$_3$As$_3$ structure. In
order to make connections with the 233-compound, we have also
performed calculations on a hypothetical crystal structure of \k133\
which is initially with space group P6$_3$/m (No. 176) but without
the rotation of the [CrAs]$_6$ sub-nanotubes (FIG. \ref{fig:geo2}).
Interestingly, this structure spontaneously turns into P$\bar{6}$m2
(No. 187) after optimization. So the loss of symmetric center is not
due to the distribution of K atoms between $z=0.25$ and $z=0.75$
planes, but rather relates to the rotation of the [CrAs]$_6$
sub-nanotubes.

Table \ref{tab:energy} lists the total energies of \k133\ with
different magnetic configurations. Unlike the K$_2$Cr$_3$As$_3$
compound, where multiple magnetic states are energetically almost
degenerate in DFT calculations, the inter-layer antiferromagnetic
(IAF) configuration, a kind of block spin magnetic order, is
apparently energetically much lower and is stable against structural
optimization. Also, it is worthy noting that both the non-collinear
phases (IOP and NCL) are energetically much higher than the
collinear phases (FM and IAF). Upon optimization, the collinear
phases are energetically stable while the non-collinear ones are
not. For the lowest energy state IAF, the magnetic moment on each Cr
is calculated to be $\sim$0.77 $\mu_B$, in good agreement with the
experimental observation of 0.68 $\mu_B$\cite{Cao133}.


\begin{table*}
 \caption{Total energies of \k133\ with different magnetic configurations. Column Expt, Opt and P$\bar{6}$m2 list the results with experimental lattice constants and coordinates, the optimized lattice constants and coordinates, and the hypothetical structure after optimization, respectively. Please be noted that the hypothetical structure was initially centrosymmetric P6$_3$/m (No. 176), but spontaneously turned into P$\bar{6}$m2 (No. 187) after optimization.\label{tab:energy}}
 \begin{tabular}{c|c|c|c|c|c|c||c|c|c|c|c}
 \hline\hline
     &   & FM & IAF & IOP & NCL & NM & BIAF$^c$ & UUD$^c$ & FM$^c$ & IAF$^c$ & NM$^c$\\
 \hline
     & $a$ ($c$) & \multicolumn{5}{c||}{9.0886 (4.1801)} & \multicolumn{5}{c}{9.0886 (4.1801)} \\
 Exp & $\Delta E$ (meV) & -41.3 (-43.9) & -121.7 (-119.4) & -22.5 & -16.0 & 0.0 & -20.7 & -23.4 & -50.7 & -129.7 & 0.0\\
     & $m_{\mathrm{Cr}}$ ($\mu_B$) & 0.36 & 0.77 & 0.51 & 0.31 & 0.00 & 0.44 & 0.40 & 0.37 & 0.78 & 0.00\\
 \hline
     & $a$ ($c$) & 9.3442 (4.0862) & 9.3419 (4.0974) & \multicolumn{2}{c|}{to NM} & 9.3447 (4.0826) \\
 Opt & $\Delta E$ (meV) & -0.8 (-6.0) & -20.4 & \multicolumn{2}{c|}{to NM} & 0.0 \\
     & $m_{\mathrm{Cr}}$ ($\mu_B$) & 0.14 & 0.29 & \multicolumn{2}{c|}{to NM} & 0.0\\
 \hline
     & $a$ ($c$) & 9.7091 (4.0848) & 9.7067 (4.0904) & \multicolumn{2}{c|}{to NM} & 9.7109 (4.0826) \\
 P$\bar{6}$m2 & $\Delta E$ (meV/Cr) & -1.6 (-4.1) & -7.0 & \multicolumn{2}{c|}{to NM} & 0.0 \\
     & $m_{\mathrm{Cr}}$ ($\mu_B$) & 0.1 & 0.24(-0.29) & \multicolumn{2}{c|}{to NM} & 0.00 \\
 \hline\hline

 \end{tabular}
\end{table*}

  \begin{figure}[htpb]
    \subfigure[]{
      \rotatebox{270}{\includegraphics[width=6cm]{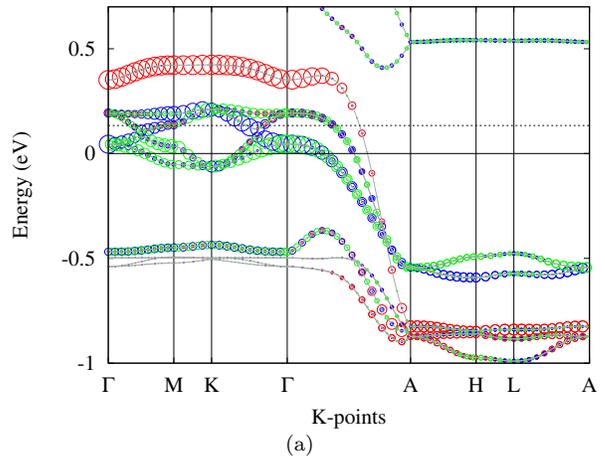}}
      \label{fig:pmbs}}
    \subfigure[]{
      \rotatebox{270}{\includegraphics[width=6cm]{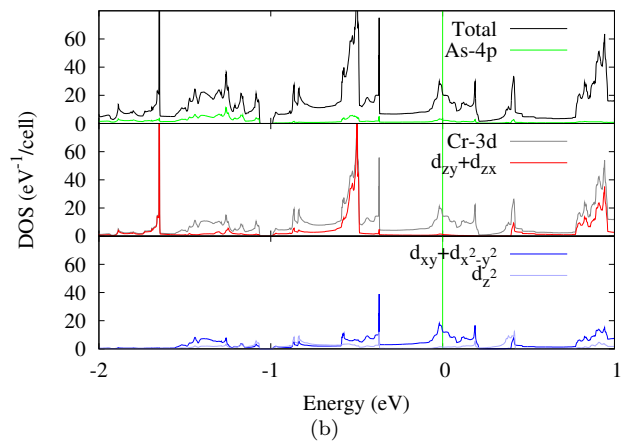}}
      \label{fig:pmdos}}
    \caption{(a) Band structure and (b) DOS of non-magnetic \k133. In panel (a), the size of the red/blue/green circles are proportional to the orbital weight of the Cr-3d$_{z^2}$/-3d$_{x^2-y^2}$/-3d$_{xy}$, respectively. The dashed horizontal line indicates the Fermi level corresponding to doping the system with 1 electron/formula, i.e. K$_2$Cr$_3$As$_3$.\label{fig:pmbsdos}}
  \end{figure}

In order to understand the formation of magnetic moments, we first
examine the electronic structure of \k133 in the non-magnetic phase.
 As shown in FIG. \ref{fig:pmbsdos} (a), the band structure of
\k133\ near the Fermi level is quite similar to that of
K$_2$Cr$_3$As$_3$, including the orbital characters, but the
degeneracies at $\Gamma$ and K are completely different. Indeed, the
degeneracies at K of \k133\ and the double degeneracy of the bands
around $E_F$-0.5 eV at $k_z$=0.5 (A-H-L-A) are direct consequences
of the P6$_3$/m symmetry which is broken in K$_2$Cr$_3$As$_3$.
Furthermore, the reduction of 50\% potassium atoms causes hole
doping to the system, resulting in a shift of the Fermi level. As a
result, the number of bands cross the Fermi level becomes five,
among which three are quasi-1D and two are 3D. The DOS and the
orbital contributions are shown in FIG. \ref{fig:pmdos}. The
quasi-1D signature van Hove singularities are apparent around -1.6
eV, -0.5 eV, -0.4 eV, 0.5 eV relative to $E_F$. The DOS around $E_F$
is dominated by the Cr-3d orbitals, in particular, Cr-3d$_{xy}$ and
3d$_{x^2-y^2}$ orbitals. The contribution of Cr-3d$_{z^2}$ orbital
is small, but non-negligible since the $\alpha$-band is almost
exclusively contributed by Cr-3d$_{z^2}$ (FIG. \ref{fig:pmbs}). The
Fermi surface (FS) of \k133 constitutes of three quasi-1D sheets and
two 3D sheets as shown in the Supplemental Material (SM). If the
system is doped with one electron per formula (i.e.
K$_2$Cr$_3$As$_3$), the FS sheets will be reduced to two quasi-1D
and one 3D, as in the original K$_2$Cr$_3$As$_3$ compound
\cite{PhysRevX.5.011013}. Of course, the shape of the 3D FS sheet will be
different from the one in the original K$_2$Cr$_3$As$_3$ due to the
local structure difference.



\begin{figure}[htpb]
    \subfigure[]{
      \rotatebox{270}{\includegraphics[width=6cm]{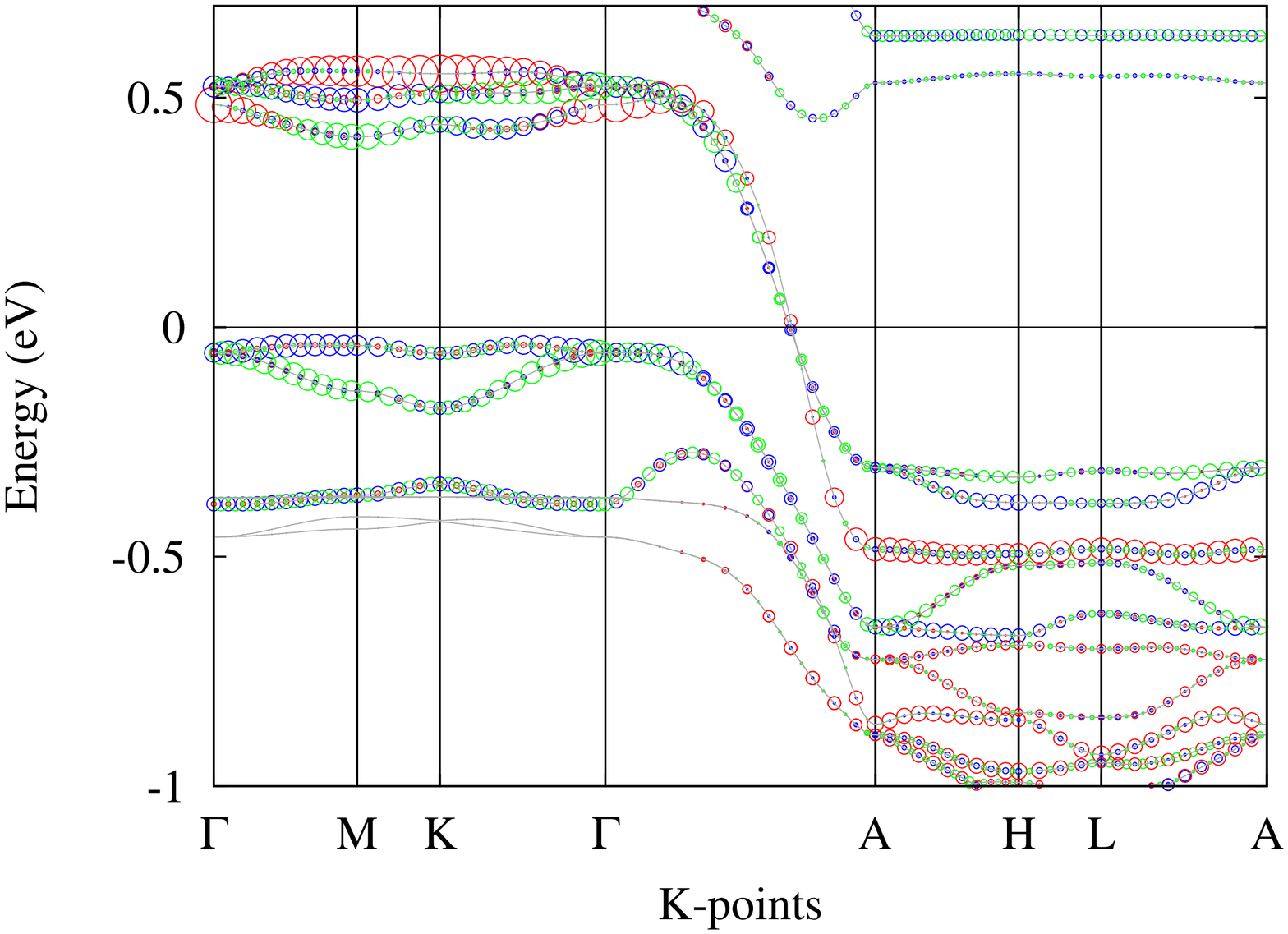}}
      \label{fig:iafbs}}
    \subfigure[]{
      \rotatebox{270}{\includegraphics[width=6cm]{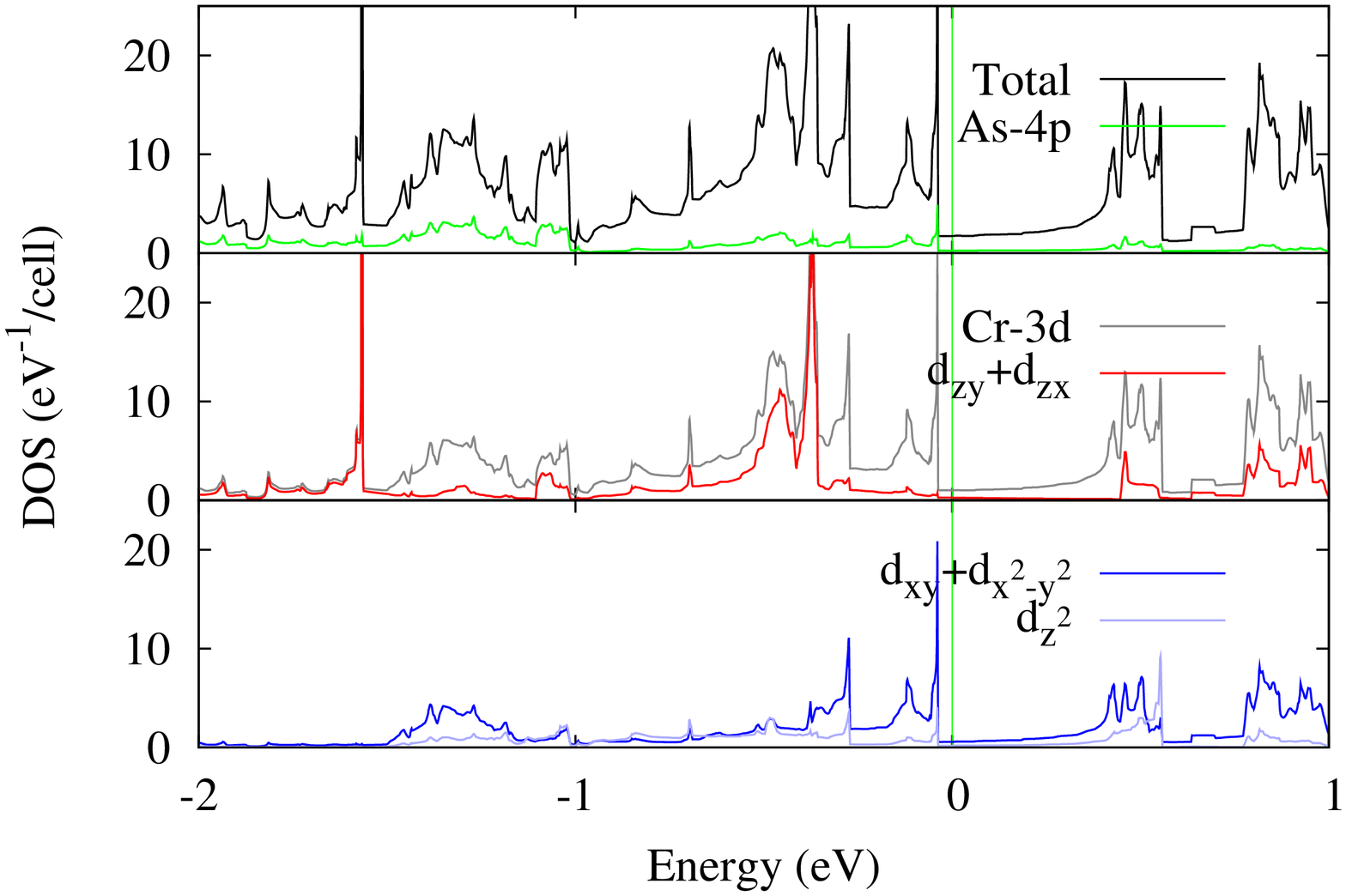}}
      \label{fig:iafdos}}
    \caption{(a) Band structure and (b) DOS of \k133\ at the IAF state. Only spin up channel is shown for clarity purposes. In panel (a), the size of the red/blue/green circles are proportional to the orbital weight of the Cr-3d$_{z^2}$/-3d$_{x^2-y^2}$/-3d$_{xy}$, respectively. The dashed horizontal line indicates the Fermi level corresponding to doping the system with 1 electron/formula, i.e. K$_2$Cr$_3$As$_3$.\label{fig:iafbsdos}}
  \end{figure}

We now show the electronic structure of the lowest interlayer
antiferromagnetic phase (IAF) in Figure \ref{fig:iafbsdos}. The
orbital characteristics of bands near $E_F$ remains the same as
those in the NM state. The four Cr-3d$_{x^2-y^2}$/Cr-3d$_{xy}$ bands
near $E_F$ are separated by $\sim$0.4 eV in the IAF phase,
presumably due to formation of local moment on Cr; whereas these
bands are very close to each other in the NM phase. As a result, two
of these bands, who formed the 3D FS sheets in the NM phase, become
fully occupied, leaving us only three quasi-1D sheets in the IAF
phase. It is worthy noting that the band around $E_F$-0.03 eV are
very flat, yielding strong quasi-1D van Hove singularity in the DOS
plot. These electrons may give rise to activation-like behavior in
resistivity\cite{Cao133}.

{\it Local Moments and Magnetic Frustration.} The above DFT
calculations suggest the existence of local moments in Cr$_3$As$_3$,
given the facts that the calculated magnetic moment $\sim 0.77
\mu_B$ is close to the saturated value of $S=1/2$ isolated local
moment and quantum frustrations are not considered so far. The local
moment phase is also hinted in the non-magnetic band structure FIG.
\ref{fig:pmbsdos} (a), where two upper degenerate bands along the
$\Gamma\rightarrow A$ direction intersect the Fermi energy at the
Fermi point $k_{F_z}\sim 0.56 \pi$. This value is very close to
$\pi/2$, indicating the nearly half-filling case of the
corresponding molecular orbital band. In 1D, any repulsive Coulomb
interaction will induce the $4k_{F_z}$ Umkllamp scattering at
half-filling, resulting in a finite charge gap. Therefore, a Mott
transition in this band is naturally expected in the corresponding
molecular orbital band\cite{Zhong15}. In fact, compared with the
band structure of K$_2$Cr$_3$As$_3$, both the band narrowing effect
and the electron correlations are enhanced due to the reduced
dimensionality in KCr$_3$As$_3$, providing more evidence for a Q1D
Mott phase via orbital selective Mott
transitions\cite{Liebsch04,Medici09} as in iron
pnictides\cite{Yu12,Valenzuela13,Rincon14}.

We are then led to consider a local moment at each Cr site described
by a spin operator, ${\bf S}_{\bf r}$, with the magnitude $S=1/2$.
Assuming that the coupling between itinerant electrons and local
moments is relatively weak, we consider the Hamiltonian of these
localized spins as
\begin{eqnarray}
{\cal H}_{\rm spin}&=&J_{1}\sum_{n,I,I'}{\bf S}_{n,I}\cdot {\bf
S}_{n,I'}
+{\tilde J}_{1}\sum_{n,{\bar I},{\bar I'}}{\bf S}_{n,\bar I}\cdot {\bf S}_{n,\bar I'}\\
&+&J_{2}\sum_{n,I,\bar I'}{\bf S}_{n,I}\cdot {\bf S}_{n,\bar I'}
+{\tilde J}_{2}\sum_{n,I,\bar I'}{\bf S}_{n,I}\cdot {\bf
S}_{n+1,\bar I'}.\nonumber
\end{eqnarray}
Where, the first two terms indicate the intra-layer nearest-neighbor
couplings in the two different triangles, while the last two terms
the inter-layer nearest-neighbor couplings between the two triangles within one unit cell and across two unit cells, respectively.

The spin model is an extended variant of twisted spin tubes
previously proposed for the three leg systems
[(CuCl$_2$tachH)$_3$]Cl$_2$ and
CsCrF$_4$.\cite{Seeber04,Schnack04,Manaka09,Sakai10}  The magnetic
frustrations will arise not only due to the intralayer spin exchanges
but also the interlayer or inter-unit-cell spin exchanges if they
are all antiferromagnetic. Concerning the present KCr$_3$As$_3$
compound, we can assume $J_{1}={\tilde J}_{1}$, $J_{2}={\tilde
J}_{2}$. Their values are estimated by fitting the corresponding
energies in different magnetic structures with the moment
per Cr atom being fixed at 0.77 $\mu_B$. The best fitting suggests $J_{1}=-9.9$
meV and $J_{2}=18.5$ meV. It is worthy noting here that we have 
also determined the strength of the inter-tube interactions $J_t$ for the IAF 
state to be $<0.1$ meV by comparing the DFT total energies of inter-tube-FM 
and inter-tube-AFM states. Thus the KCr$_3$As$_3$ system is a truly remakable
example of nearly one-dimensional spin chain, explaining why the IAF long range order was
not observed in experiments. The ferromagnetic $J_{1}$ implies the
suppression of the magnetic frustration in this 133-compound. In
this case, the twisted tube is basically a rhombus lattice without
frustration due to the bipartite nature of the lattice. A large
block-spin $S_{\triangle}=3/2$ state is then expected for each Cr
triangle, involving the four polarization components:
$|\frac{3}{2}\rangle=|\uparrow\uparrow\uparrow\rangle$,
$|-\frac{3}{2}\rangle=|\downarrow\downarrow\downarrow\rangle$,
$|\frac{1}{2}\rangle=\frac{1}{\sqrt 3}[
|\uparrow\uparrow\downarrow\rangle
+|\uparrow\downarrow\uparrow\rangle
+|\downarrow\uparrow\uparrow\rangle]$,
$|-\frac{1}{2}\rangle=\frac{1}{\sqrt 3}[
|\uparrow\downarrow\downarrow\rangle
+|\downarrow\uparrow\downarrow\rangle
+|\downarrow\downarrow\uparrow\rangle]$. By introducing a spin-3/2
operator ${\cal T}$ acting on these $S_{\triangle}=3/2$ states, we
obtain the effective Hamiltonian
\begin{eqnarray}
{\cal H}_{S_{\triangle}=3/2}&=&\sum_{n}[K_1{\cal T}_{n}\cdot {\cal
T}_{n+1}+K_2({\cal T}_{n}\cdot {\cal T}_{n+1})^2]
\end{eqnarray}
with $K_1=2J_2/3 +J^2_{2}/(18|J_1|)$ and
$K_2=-J^2_2/(54|J_1|)$\cite{Fouet06}. The low temperature property
is dominated by the $K_1$-term. With quantum fluctuations the
interlayer antiferromagnetic order is only quasi-long-ranged, in the
sense that the spin-spin correlation function exhibits a power-law
behavior and the spin excitation spectrum is gapless.

This magnetic ground state remains stable in a wider parameter
regime when $J_{1}$ moves toward the antiferromagnetic side.
However, when $J_{1}$ is much larger than $J_{2}$, the ground state
will be changed due to strong frustrations. In the limiting case,
each triangle can be solved independently, giving rise to the doubly
degenerated ground states involving $
|\uparrow_R\rangle=\frac{1}{\sqrt 3}[
|\uparrow\uparrow\downarrow\rangle
+\omega|\uparrow\downarrow\uparrow\rangle
+\omega^{-1}|\downarrow\uparrow\uparrow\rangle]$,
$|\uparrow_L\rangle=\frac{1}{\sqrt 3}[
|\uparrow\uparrow\downarrow\rangle
+\omega^{-1}|\uparrow\downarrow\uparrow\rangle
+\omega|\downarrow\uparrow\uparrow\rangle]$ (and the states with
$|\uparrow\rangle \Leftrightarrow |\downarrow\rangle$). Noted that
these lower block-spin $S_{\triangle}=1/2$ states are chiral in
nature, associated with the $C_3$ rotation $\omega=e^{i2\pi/3}$. To
the lowest order in $J_2$, the effective spin model describing these
$S_{\triangle}=1/2$ spin excitations is given by
\cite{Kawano97,Luscher04}
\begin{eqnarray}
{\cal H}_{S_{\triangle}=1/2}&=&\frac{2J_{2}}{3}\sum_{n}{\bf
T}_{n}\cdot {\bf
T}_{n+1}[1+2(\omega^2\tau^{+}_{n}\tau^{-}_{n+1}+h.c.)],\nonumber
\end{eqnarray}
with ${\bf T}$ being a spin-1/2 operator acting on the two different
spin polarization components $\sigma=\uparrow,\downarrow$, while
$\tau^{\pm}$ on the chirality degrees of freedom, namely, $
\tau^{+}|\sigma L\rangle=0, \tau^{-}|\sigma R\rangle=|L\rangle,
\tau^{+}|\sigma R\rangle=|L\rangle, \tau^{-}|\sigma R\rangle=0$.
According to the early numerical studies, this model has a
disordered ground state, i.e., the dimerized ground state exhibiting
gaps in both spin and chirality
excitations.\cite{Kawano97,Luscher04}

It is well-known from the celebrated Lieb-Schultz-Mattis
theorem\cite{Lieb61} that half-spin chains have either a degenerated
ground state or gapless spin spectrum. Interestingly, with
increasing ratio $\alpha=J_{2}/J_{1}$, both situations could be
realized in the present material. The quantum phase transition from
gapless to gapful phases takes place around $\alpha_c\sim
1.22$\cite{Okunishi05,Fouet06}.

{\it Summary and discussions. } Our DFT calculations show that the
ground state of the \k133 compound is metallic and extremely Q1D
dimensional compared with its superconducting 233-counterpart.
Moreover, it exhibits an interlayer antiferromagnetic order with a
large block-spin $3\times 0.77\mu_B\sim 2.3\mu_B$ in each Cr
triangles. By fitting the twisted  $J_{1}$-$J_{2}$ spin tube model,
we find that the basic magnetic property can be actually captured by
the quantum antiferromagnetic Heisenberg chain with block-spin
$S_{\triangle}=3/2$ defined in Eq.(2). The absence of
superconductivity is likely due to the strong competition tendency
of the quasi antiferromagnetic order in this 133-compound.

We should remark that due to the reduced dimensionality, a static
long-range magnetic order is unstable against to enhanced quantum or
thermal fluctuations, and strongly influenced by the disorder
effect. So it is not very strange that the spin glass behavior,
instead of the static magnetic order, is observed in the
polycrystalline samples\cite{Cao133}. We also remark here that
although the ground state itself is gapless, the lower spin
$S_{\triangle}=1/2$ sector may still play some role at finite
temperatures, exhibiting a crossover-to-spin-gap behavior as in the
spin tube material [(CuCl$_2$tachH)$_3$]Cl$_2$\cite{Ivanov10}.

An intriguing situation with a large intralayer antiferromanetic
coupling $J_{1}$ is expected, hopefully by applying the planar
pressure while keeping the symmetry of charge distributions among
the two conjugated triangles. Such situation should be contrasted to
the 233-compound, where the more $K$ concentration induces the
non-centrosymmetric structural distortion and asymmetric charge
distributions in the two conjugated triangles. Because
superconductivity may be also induced from the spin-gaped phase upon
doping\cite{Rice93} and given the recent pressure experiments on the
K$_2$Cr$_3$As$_3$\cite{Kong15,Wang15}, we expect that the disordered
phase with both spin and chirality gaps can be realized in further
coming pressure experiments on KCr$_3$As$_3$ and the related
compounds.

\begin{acknowledgments}
The authors would like to thank Guanghan Cao, Jiangping Hu, Zhong-Yi
Lu, Qimiao Si, Zhuan Xu, Fuchun Zhang, and Yi Zhou for inspiring
discussions. This work has been supported by the NSFC (Nos.
11274006, 11274084, and 11304071), National Basic Research Program
(No. 2014CB648400) and the NSF of Zhejiang Province (No.
LR12A04003). All calculations were performed at the High Performance
Computing Center of Hangzhou Normal University College of Science.
\end{acknowledgments}

\bibliography{K133}

\newpage
\clearpage

\renewcommand{\theequation}{S\arabic{equation}}
\setcounter{equation}{0}
\renewcommand{\thetable}{S\arabic{table}}
\setcounter{table}{0}
\renewcommand{\thefigure}{S\arabic{figure}}
\setcounter{figure}{0}

\section*{SUPPLEMENTARY MATERIAL}

\section*{Reduced Dimensionality and Magnetic Frustration in KCr$_3$As$_3$}

by: Chao Cao, Hao Jiang, Xiao-Yong Feng, and Jianhui Dai

\vskip 1.0 cm


 \begin{figure}[htpb]
    \subfigure[Prestine \k133]{
      \includegraphics[width=8cm]{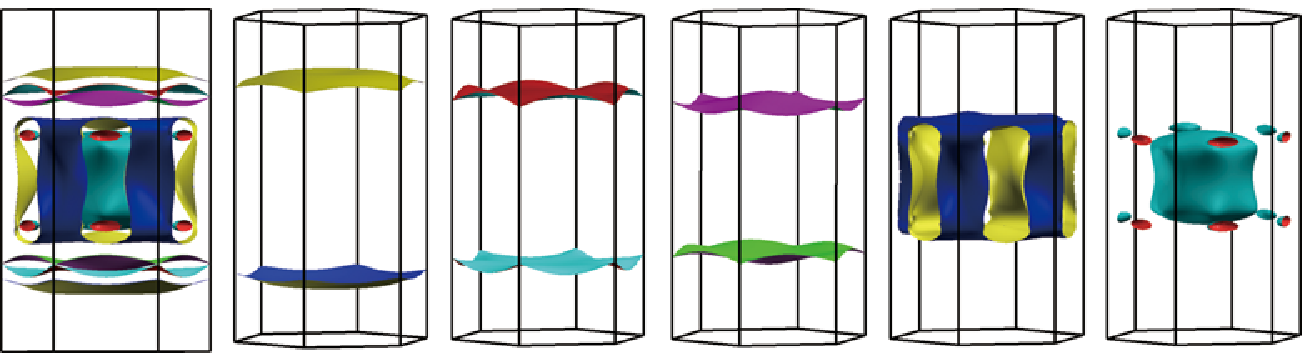}}
    \subfigure[electron doped \k133]{
      \includegraphics[width=8cm]{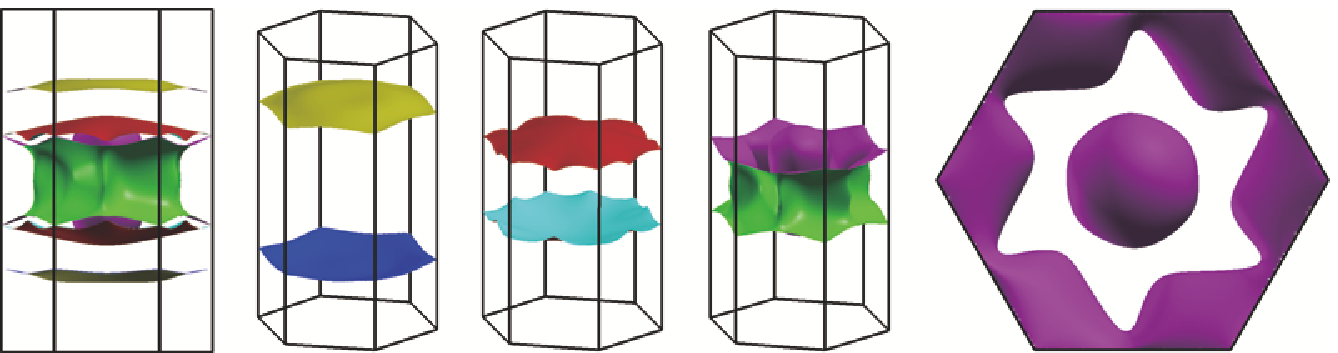}}

    \caption{Fermi surfaces of (a) prestine \k133\ and (b) electron-doped \k133. \label{fig:pmfs}}
  \end{figure}

The Fermi surfaces are reconstructed by fitting the
first-principles band structure to a 10-orbital tight-binding
Hamiltonian using the maximally localized Wannier functions (MLWFs).
Figure \ref{fig:pmfs}(a) shows the Fermi surface sheets of \k133.
Three quasi-1D sheets and two 3D sheets constitute the Fermi surface
of prestine \k133 as expected. If the system is doped with 1e per
formula (i.e. K$_2$Cr$_3$As$_3$), the FS sheets will be reduced to
two quasi-1D and one 3D, as in the original K$_2$Cr$_3$As$_3$
compound \cite{Jiang14}. However, the shape of the 3D FS sheet will be
completely different from the one in the original K$_2$Cr$_3$As$_3$
due to the local structure differences.

\end{document}